# In-Gap Quasiparticle Excitations Induced by Non-Magnetic Cu Impurities in Na(Fe$_{0.96}$Co$_{0.03}$Cu$_{0.01}$)As Revealed by Scanning Tunneling Spectroscopy

Huan Yang[1,†], Zhenyu Wang[2,†], Delong Fang[1,†], Qiang Deng[1], Qiang-Hua Wang[1],

Yuan-Yuan Xiang[1], Yang Yang[1], Hai-Hu Wen[1,*]

[1] Center for Superconducting Physics and Materials, National Laboratory of Solid State Microstructures and Department of Physics, National Center of Microstructures and Quantum Manipulation, Nanjing University, Nanjing 210093, China

[2] National Laboratory for Superconductivity, Institute of Physics and National Laboratory for Condensed Matter Physics, Chinese Academy of Sciences, Beijing 100190, China

[†] These authors contributed equally to this work

[*] To whom correspondence should be addressed. E-mail: hhwen@nju.edu.cn

**The pairing mechanism in the iron pnictides remains unresolved yet. One of the central issues is the structure of the superconducting order parameter which classifies the community into two different and highly disputed camps. On one hand the picture of pairing based on the magnetic origin predicts a sign reversal gap on the electron and hole Fermi pockets, leading to the S$^\pm$ pairing. On the other hand, a more conventional S$^{++}$ pairing gap was suggested based on the**

**phonon or orbital fluctuation mediated pairing. In the superconducting state, the impurities may generate a unique pattern of local density of states in space and energy, which are regarded as the *fingerprints* for checking the structure of the pairing gap. In this study, we successfully identified the nonmagnetic and magnetic impurities in Na(Fe$_{0.97-x}$Co$_{0.03}$T$_x$)As (T=Cu, Mn) and investigated the spatial resolved scanning tunneling spectroscopy. We present clear evidence of the in-gap quasiparticle states induced by the nonmagnetic Cu impurities, giving decisive evidence of the S$^{\pm}$ pairing. This is corroborated by the consistency between the experimental data and the first-principles calculations based on the S$^{\pm}$ pairing gap with a scalar scattering potential.**

Since the discovery of high temperature superconductivity in the iron pnictides and chalcogenides in early 2008 (*1*), the pairing mechanism remains unsolved yet. It was proposed that the binding force between the two electrons of the Cooper pair might be established through the magnetic origin, either by exchanging the anti-ferromagnetic spin fluctuations or through local magnetic interactions, and consequently the electron and hole pockets should have s-wave gap with opposite signs (the S$^{\pm}$ pairing model) (*2,3*). As far as we know, there are no experiments that give direct evidence to support the S$^{\pm}$ model. This gap structure gets support from the scanning tunneling spectroscopy (STS) and the quasi-particle interference (QPI) measurements in Fe(Se,Te) (*4*), in which a unique **Q$_2$** interference spot was observed reflecting the quasiparticles (QP) induced by the non-magnetic impurities within the S$^{\pm}$ pairing model. Another indirect evidence was given by the observation of the resonance peak of the imaginary part of the spin susceptibility at $(\pi,\pi)$ in the inelastic neutron scattering experiment ([5]). In the case of $S^{\pm}$ pairing, although there is a sign reversal of the gap, because the individual Fermi surface is fully gapped, one may not be able to detect it through the phase-sensitive experiment as conducted in the cuprate superconductors ([6]). Alternatively, one can probably detect this unique pairing model through the impurity effect. In the superconducting state, the impurities may induce the QP scattering, break Cooper pairs and generate a unique pattern of the local density of states (LDOS) of QPs. This pattern is strongly dependent on the pairing gap structure and the characteristics of the impurities ([7,8]). According to the Anderson's theorem ([9]) and the subsequent calculations ([7,8,10]), the Cooper pairs with singlet pairing can survive in the presence of non-magnetic impurities, while the magnetic impurities are very detrimental to superconductivity yielding a unique pattern of the QP-LDOS. This impurity state, after the pioneer work of Yu-Shiba-Rusinov ([10]), has been well proved in the conventional superconductor Nb by the STS measurements using Mn adatoms as the magnetic scattering centers ([11]). In the cuprate superconductors, when doping Zn atoms (the usually believed non-magnetic impurities) to the Cu sites, the superconductivity is suppressed greatly and a strong peak of the LDOS is observed near bias zero energy at the Zn impurity, which suggests an unconventional pairing symmetry and an anomalous normal state ([12]). Theoretically it was argued that the $S^{\pm}$ type pairing should be also fragile to the non-magnetic impurities when the scattering potential ($V_{imp}$) is moderate ($V_{imp} \geq 1eV$) ([13-17]), namely the in-gap QP excitations should be observed. However, some experiments show a slow suppression rate of superconducting transition temperature to impurity scattering, which seems violate the Anderson's theorem and leads to the conjecture of the $S^{++}$ pairing (*13*). Furthermore, a picture of orbital fluctuation mediated superconductivity was proposed based on the analysis of five orbital Hubbard-Holstein model, which predicts a $S^{++}$ pairing (*18*). In fact, the spatial and energy dependent LDOS induced by the impurity scattering are very essential for having a formal check on the pairing model of $S^{\pm}$ or $S^{++}$.

In previous STM studies, impurities have been successfully located in iron pnictide superconductors and many interesting features have been reported (*19-25*). However, it remains to be a great challenge to figure out which impurities are magnetic or nonmagnetic. It is highly desired to locate a *well-defined quantum impurity, proof by experiment whether they are magnetic or nonmagnetic and investigate the related LDOS at and nearby the impurity*. In our previous study, we identified the Co impurities and found an invisible spatial evolution of the STS crossing a Co impurity in Na(Fe$_{1-x}$Co$_x$)As(*24*), which is explained in the picture that the Co impurity may give a weak and extended scattering potential, induce mainly the intra-pocket scattering and thus cannot break the Cooper pairs assuming that the pairing is established by the inter-pocket scattering. In the present study we show the STS data obtained on Na(Fe$_{0.96}$Co$_{0.03}$T$_{0.01}$)As single crystals with T = Mn ($T_c \approx$ 12.8 K) and Cu ($T_c \approx$ 16.0 K). Our DC magnetization measurements indicate that the Mn dopants will contribute magnetic, while the Cu dopants give rise to non- or very weak magnetic impurities (*26*). However, strong in-gap quasiparticle DOS are clearly observed for either Cu or Mn impurities. Our results give decisive evidence for the sign reversal gap structure, i.e., the $S^{\pm}$ pairing in the present samples.

Fig.1**A** and **B** present the topographic images of the cleaved surfaces with the orientation (001) of Cu and Mn doped single crystals Na(Fe$_{0.96}$Co$_{0.03}$T$_{0.01}$)As (T = Cu and Mn).

In the Na(Fe$_{1-x}$Co$_x$)As system, the optimal doping is achieved around x = 0.03 with superconducting transition temperature T$_c$ = 20.5 K. Further doping of Cu and Mn to the Fe site is achieved starting from the status of this pristine sample. In the Cu doped samples, in addition to some 3×2 blocks with light bright color, as we discovered in the pristine Co-doped samples ([24]), one can clearly see some extra spots with a dumbbell shape and much stronger brightness. The former are corresponding to the Co-doped sites, as we described clearly in previous paper ([24]), the latter are the impurity sites induced by the Cu atoms. The two different orientations of these dumbbells for the Cu impurities have the same origin as the Co impurities, mainly been induced by the selective substitution to the Fe site which is surrounded by six Na atoms on the top layer. We illustrated this in the Supplementary Materials. A scan with much higher pixel was taken and the data are shown for the region near a Cu impurity in the inset of Fig. 1**A**. A close scrutiny finds that the brightness of the dumbbell decays actually in a scale of about 15-20 angstrom, being much larger than the Co impurity, which will be further corroborated by the spatial evolution of the LDOS. In the Mn doped samples, since the image was taken with a lower voltage (8 meV), the Co sites are barely visible. However, we clearly see some spots with a dark-bright crossing pattern. As amplified in the inset of Fig.1 **B**, the dark bar seems to be slightly longer than the bright wing. Clearly the Mn impurity gives a very different pattern of image compared with the Cu impurity. Qualitatively one may categorize the Co and Cu impurities as the same type, namely non- or weak magnetic, since they have quite similar shapes. This will be corroborated by the DC magnetization measurements illustrated below. The reason for the different patterns of image between Cu (also Co) and Mn dopants remains unresolved yet and is highly desired for further theoretical input.

Although the topographic images of the Cu and Mn impurities are very different, they suppress the superconductivity in a similar way. In Fig.1 **C**, we present the temperature dependence of the resistivity of the pristine samples Na(Fe$_{0.97}$Co$_{0.03}$)As with T$_c$ =20.5 K, and the Mn and Cu doped samples with doped concentration of 1%, 2% and 3%. One can see that the superconducting transition temperature drops down with the increase of doping as well as the residual resistivity. Interestingly, the suppression rate of $\Delta T_c/\Delta x$ (as shown in Fig.1 **D**) is quite close to each other for the two doping ways. For evaluating the magnetic moments induced by these dopants, we present in Fig.1 **E** the temperature dependence of DC magnetization measured at $\mu_0 H$ = 1 T for several typical samples. The magnetization hysteresis loops measured on these samples clearly show a reversible, roughly linear and paramagnetic behavior, with the absence of any long range of magnetic order above 2 K. Assuming that the magnetization can be written in the Curie-Weiss law, $\chi = M/H = \chi_0 + C_0/(T+T_\theta)$, here the first term comes from the Pauli paramagnetism which is proportional to the DOS at the Fermi energy, the second term is contributed by the paramagnetism of the local magnetic moments carried by the Fe and the dopants (Cu, Mn and Co). For the pristine sample (3% Co doped), the magnetization shows a very weak up-turn in the low temperature region indicating a weak magnetic moments per Fe. With the doping of Cu, no enhancement of the up-turn at low temperatures is observed. This is in sharp contrast with the Mn doped samples, one can see that the upturn at low temperatures is clearly getting enhanced with the increase of the doped Mn concentration. In order to get a quantitative assessment on the magnetic moments, we fit the data below 40 K for all samples based on the Curie-Weiss law and present the results in Fig.1 **F**. The detailed fitting process and results are given in the Supplementary Materials. It is clear that Cu-doping seems to lower down the averaged magnetic moments, while Mn-doping gives rise to a continuing increase of the local magnetic moments. Since the measured average magnetic moment after doping Cu is even slightly lower than the pristine sample, this suggests that the Cu impurity may diminish the local magnetic moments. One possible picture to interpret this is that the Cu dopant may have a full shell of electrons with the ionic state of $Cu^{1+}$ as predicted by the first-principles calculations (*27*) and the density functional calculation (*28*). All these facts justify that to categorize the Cu dopants as the non- or weak magnetic impurities. For Mn dopants, they certainly play as the magnetic impurities here.

In Fig.2 **A,** we show the topographic image around the single Cu impurity. After identifying the Cu impurity and confirming that they are nonmagnetic impurities, we measure the STS by crossing one single Cu impurity with steps of half Na-Na lattice constant ($a_0/2$), and show them in Fig.2 C. A gradual evolution can be easily seen here. At a distance of about 2 nm away, the STS shows a typical superconducting one with two clear coherence peaks at about ± 5 meV. This energy gap is well consistent with the values derived from the angle resolved photo-emission (ARPES) measurements (*29*). There is a lifted height on the STS near zero bias energy even when it is far away from the impurities, this may be caused by two reasons: (1) Doping Cu can induce the in-gap QP excitations, even it is far away from the impurity site; (2) The measuring temperature is 1.7 K, which may induce some thermal broadening effect and lift the conductivity at zero bias energy. When approaching the impurity site of Cu (the center of the dumbbell shape of the image), the coherence peak on the negative biased voltage is strongly suppressed and the one on the positive side shifts to a smaller voltage. Two typical STS, measured at the Cu site and 2 nm away respectively, are presented in Fig. 2 **B**. In order to know how the Cu impurity gives the influence on the STS, we subtract the STS curves measured at different positions with that measured 2 nm away and present the results in Fig. 2 **D**, one can clearly see that the difference of the STS exhibits a huge asymmetric peak within the superconducting gap. This is very different from the case when crossing the Co site, there we didn't see any visible spatial change of the STS (*24*). We will discuss these unique in-gap states later. Furthermore, to check how strong the Cu impurity would affect the electronic properties spatially, we present the spatial mapping of the LDOS measured at different energies in Fig.3**B-F**. One can see a systematic evolution of the LDOS around the Cu impurity with a spatial scale of about 1.5-2.0 nm.

For a comparison, we also measured the STS around a Mn impurity and presented the results in Fig.4. Now one can see that, even the measurement is done about 1.2 nm away from the Mn site, a very asymmetric STS can still be observed. This asymmetry may reflect the effect given by the impurity scattering which produces the QP-DOS near the gap edge. Therefore the sharp peaks at the gap edges may be due to the merging of the superconducting coherence peak together with the impurity induced states. However, this type of asymmetric curve can be even seen at a distance far away from the Mn site, which may challenge the simple understanding mentioned above. When approaching to the Mn site, one can see that the sharp peaks near the superconducting gap become broadened and move gradually to higher energies. At the meantime, an in-gap peak appears around zero bias energy and is getting enhanced very near the Mn site. This in-gap peak is certainly derived from the QP excitations by the scattering effect of the Mn impurity, and it is no doubt that the Mn impurities are playing as the strong pair breakers. The interesting point is that the STS exhibits different evolutions along the dark bar and the bright wing on the image. The modification to the STS by the scattering effect shows up at $2a_0$ from the Mn site when measured along the dark bar, while it starts at about $a_0$ when along the bright wing. Moreover the in-gap state peak seems to locate at or very near zero bias voltage when it is measured on the atom with enhanced brightness in the image. The detailed reason is still unknown and waits for further understanding. Clearly, either the nonmagnetic Cu or the magnetic Mn impurities can induce in-gap QP excitations.

In Fig.5 we present the comparison between the experiment and the theoretical calculations for the Cu impurities. In order to have an effective comparison, we present here the data normalized by the normal state background, as shown in Fig.5 A. The detailed process of normalization is given in the Supplementary Materials. On can see that the STS exhibits an elevated bottom at zero bias voltage, this is quite natural for the impurity induced pair breaking. Interestingly, at the Cu site, the coherence peak at the positive bias seems moving inward indicating the suppression to the order parameter. In Fig.5 B, we show the difference of the STS measured at the Cu site and 2 nm away. The difference of these two STS shows a prominent peak near zero bias, with the sharp peak at a slight positive bias voltage. It is interesting to mention that in the previous experiments on conventional Nb superconductor (*11*), the magnetic impurity Mn gives a very similar effect. The only difference is that the sharp peak appears at a slightly negative voltage side there. Here we observe such peak with a non-magnetic impurity, which manifests itself that we should have a sign reversal on the superconducting gap. In order to corroborate this point, we present the calculated results in Fig.5 **C** and **D**. There are two possible scenarios to interpret the impurity induced in-gap DOS observed in our experiment. The first is a pairing state with in-phase gap on all Fermi pockets but subject to magnetic impurities, as described for Mn in the conventional Nb superconductor (*11*). This scenario can be ruled out here since no prominent magnetic signature is detected on the Cu impurities. The second is a pairing state with $S^{\pm}$ pairing on electron/hole pockets subject to nonmagnetic impurities. This is because a local impurity can scatter electrons both within and between the pockets. The inter-pocket scattering sees the sign reversal of the gap function (on electron and hole pockets), hence can induce in-gap states. In reality both intra- and inter-pocket scatterings are present, and the situation is best illustrated by a model calculation. Here we use a five-orbital band structure derived from ab initio calculations. Assuming local Coulomb and Hund's rule interactions, we performed functional renormalization group calculations to get the best pairing gap function (which turns out to have $S^{\pm}$ symmetry as expected), enabling us to write down an effective Hamiltonian describing the superconducting state. The effect of local impurity is then handled by the standard T-matrix formalism. More technical details and parameters in the calculation are described in the Supplementary Materials. The results are presented in Fig. 5**C** and **D** for an attractive impurity (i.e., with negative scattering potential $V_{imp}$ = -1.5 eV). One can see a close similarity to the experimental data. We emphasize that such a feature is a unique property of $S^{\pm}$ pairing gap. It is absent, e.g., if we assumed an $S^{++}$ pairing instead. We also notice that the asymmetry in the impurity-DOS is linked to the band structure as well as the sign of the impurity potential. The negative sign for the impurity potential is consistent with the fact that there are more d-electrons on the Cu impurity site experimentally. Finally we should mention that the energy of the in-gap states could reach zero but only at a very specific value of the impurity potential neither in the Born nor the unitary limit. Therefore the generic feature is the appearance of finite-energy in-gap states, as seen in our experiment. Our first principles calculations based on the scalar scattering potential can really capture the main features of the experimental data, this gives further support and validates the observation of in-gap states by the nonmagnetic Cu impurities. Taking into account the fact that the gaps are almost isotropic ($\Delta \approx 5$ meV) on both the hole and the electron Fermi surfaces (*29*), we can argue that the observation of the in-gap states on nonmagnetic Cu impurities here give decisive evidence for the $S^{\pm}$ pairing.

**Figures and Captions**

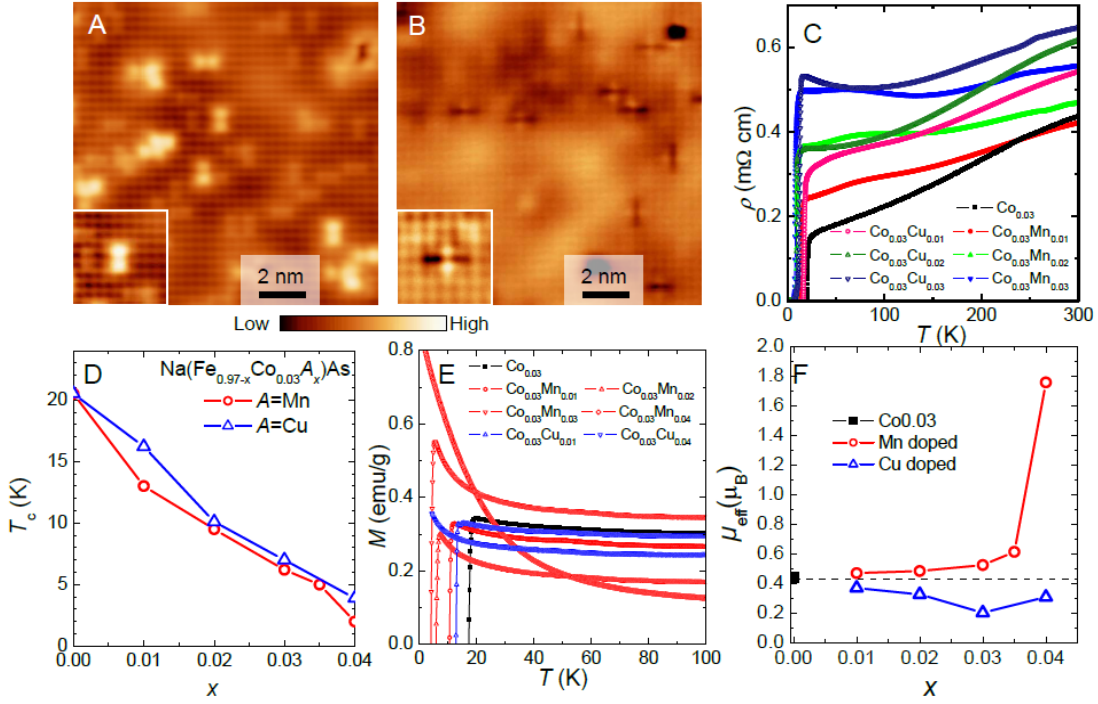

**Fig. 1.** Topographic STM images and characterizations of the Na(Fe$_{0.97}$Co$_{0.03}$)As and Na(Fe$_{0.97-x}$Co$_{0.03}$T$_x$)As single crystals (T = Cu and Mn). (**A**) and (**B**) STM images of the Cu doped and the Mn doped samples with the a bias voltage of $V_{bias}$ = 15, 8 mV and tunneling current of $I_t$ = 100, 150 pA respectively. The scale bars in (**A**) and (**B**) are 1 nm. The inset shows the rescanned data in a small area with much higher pixel, from which one can clearly see the Cu and Mn impurities. (**C**) Temperature dependence of resistivity for the pristine sample and the doped ones. The superconductivity is suppressed clearly when the residual resistivity is enhanced. (**D**) The doping dependence of the critical transition temperature T$_c$ for both the Cu and Mn doped samples. (**E**) The magnetization measured at $\mu_0 H$ = 1 T on the pristine sample (black open squares), the Cu doped samples (blue symbols) and the Mn doped samples (red symbols). One can see that the low temperature upturn gets enhanced clearly by the Mn doping, but not enhanced by the Cu doping. (**F**) The average magnetic moment per Fe site calculated by the Curie-Weiss law for the Cu and Mn doped samples. Clearly, doping Mn induces strong averaged magnetic moments, while doping Cu seems to weaken the averaged local moments.

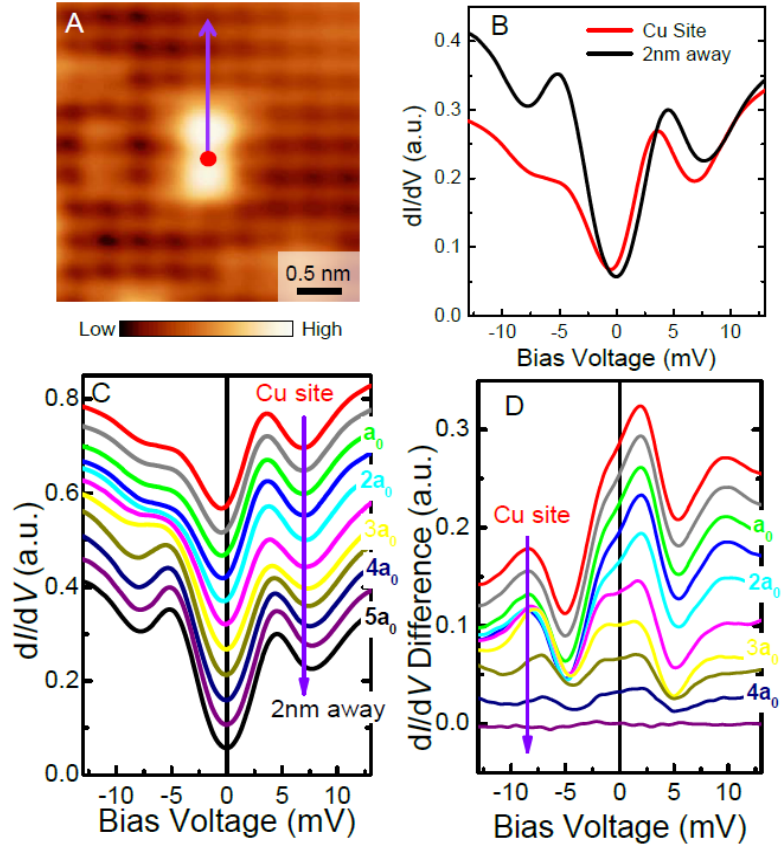

**Fig. 2.** Tunneling spectra at and near the Cu impurity. (**A**) The topographic image near a Cu impurity, the bright pattern with a dumbbell shape is corresponding to a Cu impurity. One can also see several Co impurities with lighter brightness. (**B**) The STS measured at the Cu impurity (red line) and that measured 2 nm away (dark line). (**C**) The STS measured at the Cu impurity (red line) and those measured with steps of every half Na-Na lattice constant on the surface. (**D**) The subtracted results of the STS with that measured at 2nm away from the Cu site, the colors of the lines are the same in (**C**). The STS in (**C**) and (**D**) are off-set for clarity. One can easily see the emerging in-gap states when approaching the Cu impurity.

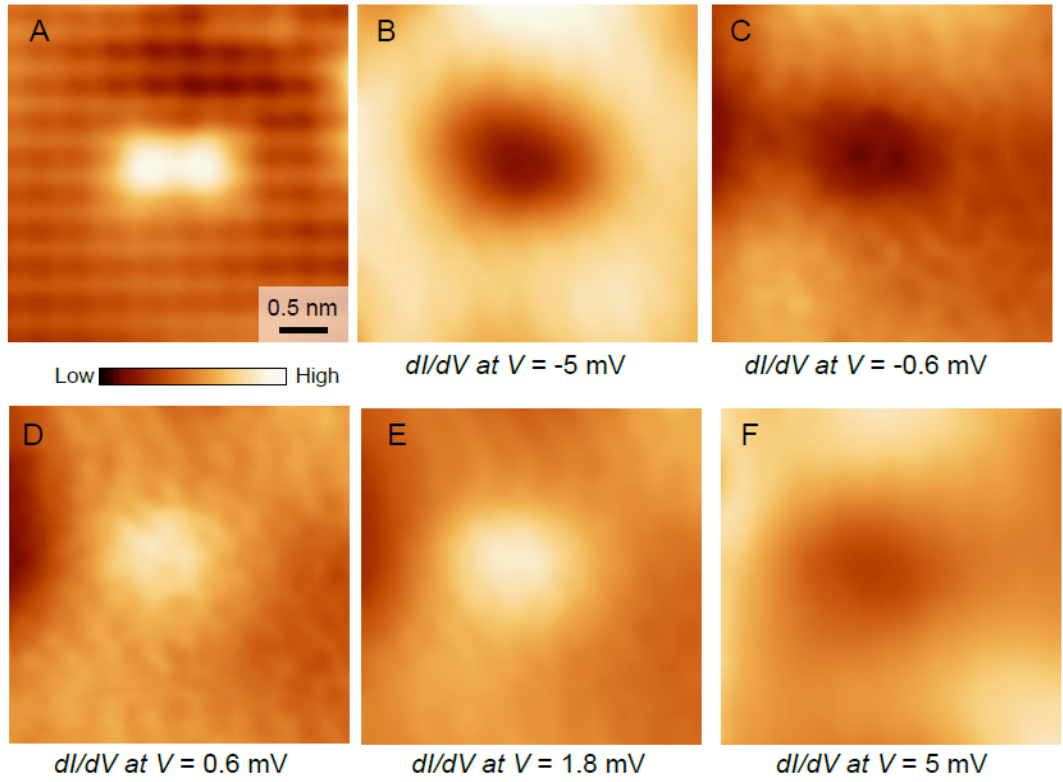

**Fig. 3.** Mapping of the LDOS measured near a Cu impurity at different bias voltages. (**A**) The topographic image of a Cu impurity, the field of view dimensions are 3.5 nm × 3.5 nm. (**B**)-(**F**) The mapping of the LDOS measured at bias voltages of -5, -0.6, 0.6, 1.8 and 5 mV, respectively. One can see that the spatial influence of a Cu impurity is about 20 Å.

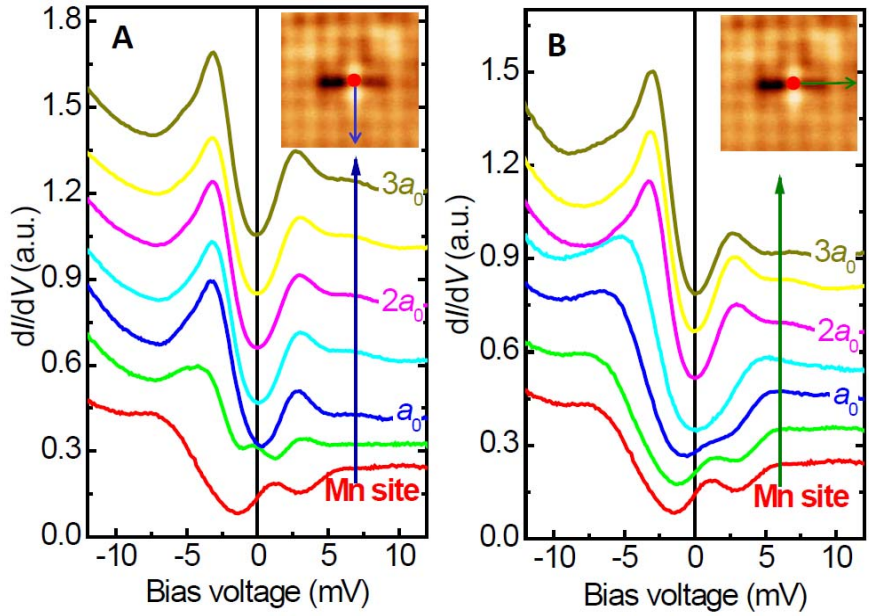

**Fig. 4.** The STS measured around a Mn impurity. (**A**), (**B**) The STS measured at the Mn site (red line) and those measured with steps of every half Na-Na lattice constant on the surface, along the line of (**A**) bright-wing and (**B**) dark-bar on the images as shown in the insets. The shape of STS exhibits a strong evolution spatially, indicating a clear presence of the in-gap states. At the Mn site and the position one step away, one can see an peak of in-gap states directly from the STS, as shown by the red and green curves.

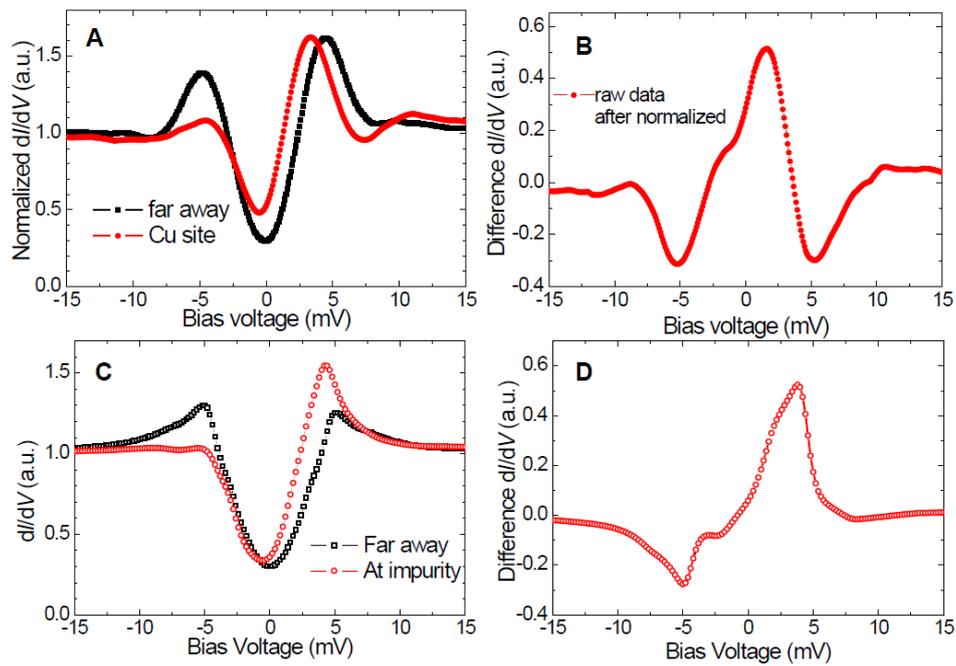

**Fig. 5.** Comparison between the experimental data and the first principles calculations. (**A**) The tunneling spectrum measured at the Cu site (red line) and that 2 nm away (black line) normalized by the normal state background. (**B**) The difference between the two curves shown in (**A**). (**C**) The STS obtained from the first-principles calculations at the scalar impurity (red filled circles) and off the impurity site (dark filled squares). (**D**) The difference between the two curves shown in (**C**). One can see that the calculation can really reproduce the main structure of the measured data. A calculation with the $S^{++}$ model was tried and could not produce the in-gap states.

# Supplementary Materials of the paper

# In-Gap Quasiparticle Excitations Induced by Non-Magnetic Cu Impurities in Na(Fe$_{0.96}$Co$_{0.03}$Cu$_{0.01}$)As Revealed by Scanning Tunneling Spectroscopy

Huan Yang[1†], Zhenyu Wang[2†], Delong Fang[1†], Qiang Deng[1], Qiang-Hua Wang[1], Yuan-Yuan Xiang[1], Yang Yang[1], Hai-Hu Wen[1*]

[1] Center for Superconducting Physics and Materials, National Laboratory of Solid State Microstructures and Department of Physics, National Center of Microstructures and Quantum Manipulation, Nanjing University, Nanjing 210093, China

[2] National Laboratory for Superconductivity, Institute of Physics and National Laboratory for Condensed Matter Physics, Chinese Academy of Sciences, Beijing 100190, China

## I. Growth and characterization of the Na(Fe$_{0.97-x}$Co$_{0.03}$T$_x$)As single crystals (T = Cu and Mn)

High quality Na(Fe$_{0.97}$Co$_{0.03}$)As and Na(Fe$_{0.97-x}$Co$_{0.03}$T$_x$)As (T = Cu and Mn) single crystals were synthesized by the self-flux method. Firstly, NaAs was prepared as the precursor. The Na (purity 99%) was cut into pieces and mixed with As powders (purity 99.99%), the mixture was put in an alumina crucible and sealed in a quartz tube in vacuum. The mixture was slowly heated to 200 °C and held for 10 hours, followed by cooling down to room temperature. Then the resultant NaAs, and Fe (purity 99.9%), Co (purity 99.9%), T (Cu or Mn) (purity 99.9%) powders were weighed with an atomic ratio of NaAs:Fe:Co:T = 4:(0.97-x):0.03:x and ground thoroughly. The mixture was pressed into a pellet and loaded into an alumina crucible, sealed in an iron tube under Ar atmosphere. Then it was placed in the furnace and heated up to 950 °C and held for 10 hours, followed by cooling down to 600°C at a rate of 3 °C/h to grow single crystals. In the preparation process, the weighing, mixing, grounding were conducted in a glove box under argon atmosphere with the $O_2$ and $H_2O$ below 0.1 PPM. The DC magnetization measurement was carried out with a SQUID-VSM-7T (Quantum Design). The resistivity measurement was done on a PPMS-16T (Quantum Design) with the standard four-probe method. The temperature dependence of DC magnetization and resistivity of the $Na(Fe_{0.97}Co_{0.03})As$ and $Na(Fe_{0.97-x}Co_{0.03}T_x)As$ (T = Cu and Mn) single crystals are shown in Fig.S1. The magnetization on the different samples show sharp superconducting transitions with $T_c \approx 20.5$ K for $Na(Fe_{0.97}Co_{0.03})As$, $T_c \approx 16$ K for $Na(Fe_{0.96}Co_{0.03}Cu_{0.01})AS$ and $T_c \approx 12.8$ K for $Na(Fe_{0.96}Co_{0.03}Mn_{0.01})As$. The resistivity data in Fig.S1 **B** show that both Cu and Mn impurity suppress the superconducting transition temperature and increase the residual resistivity.

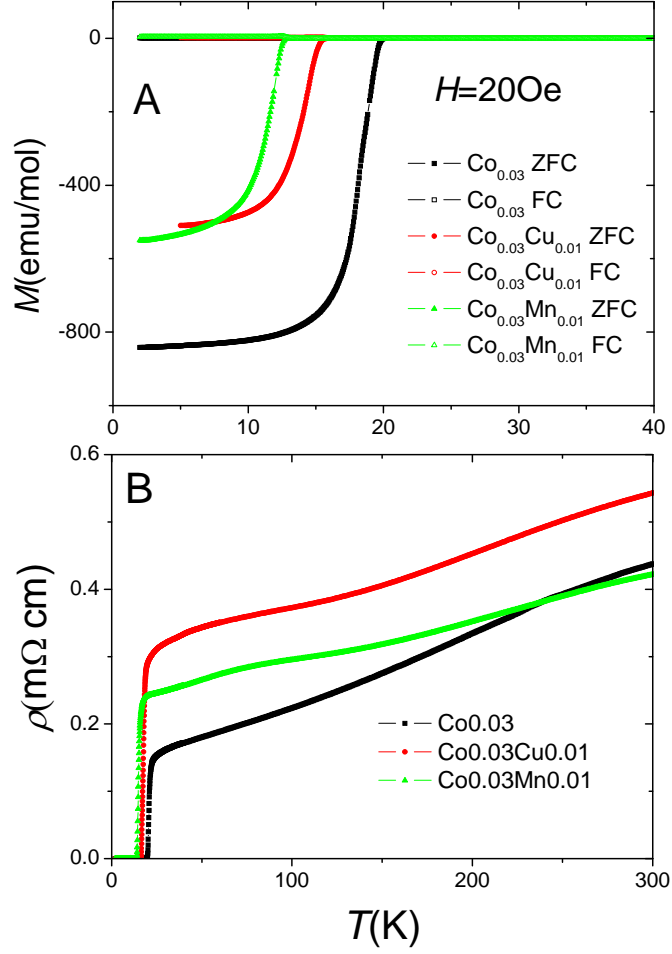

**Fig. S1.** Temperature dependence of (A) magnetization and (B) resistivity of the Na(Fe$_{0.97}$Co$_{0.03}$)As and Na(Fe$_{0.96}$Co$_{0.03}$T$_{0.01}$)As (T = Cu and Mn) single crystals. The magnetization curves were measured at a magnetic field of 20 Oe after zero-field-cooled (ZFC) and field-cooled (FC) processes.

## II. The Curie-Weiss fitting

In order to evaluate the magnetic moments induced by Cu and Mn dopants, we have done the magnetization measurement under high magnetic fields. The raw data of magnetization measurement at 1T up to 300K are shown in Fig. 1 **E**. The strong diverging of the magnetic susceptibility at low temperatures can be understood as the existence of some local magnetic moments. Assuming that the low temperature magnetization can be written in the Curie-Weiss law, $\chi = M/H = \chi_0 + C_0/(T+T_\theta)$, where $C_0 = \mu_0 \mu_{eff}^2 / 3k_B$, $\chi_0$ and $T_\theta$ are the fitting parameters, $\mu_{eff}$ is the local magnetic moment per Fe site. The first term $\chi_0$ comes from the Pauli paramagnetism of the conduction elections, which is related to the DOS at the Fermi energy. The second term $C_0/(T+T_\theta)$ is contributed by the local magnetic moments given by the Fe site (including Co, Cu, and Mn). As shown in Fig. S2, we adjust the $\chi_0$ value to get a linear function of $1/(\chi-\chi_0)$ versus $T$ in the low temperature limit. Then we fit the data with a linear function, the slope gives $1/C_0$, and the intercept provides the value of $T_\theta/C_0$. Once $C_0$ is obtained, we can get the average magnetic moment of a single Fe site (including the contribution of Fe and the dopants ). The results are shown in Fig. 1 **F**. Clearly, doping Mn induces strong local magnetic moments, while doping Cu seems to even weaken the average local moments. These facts suggest that Mn ions here play as a role of magnetic impurities, while Cu dopants act as the non- or weak magnetic impurities.

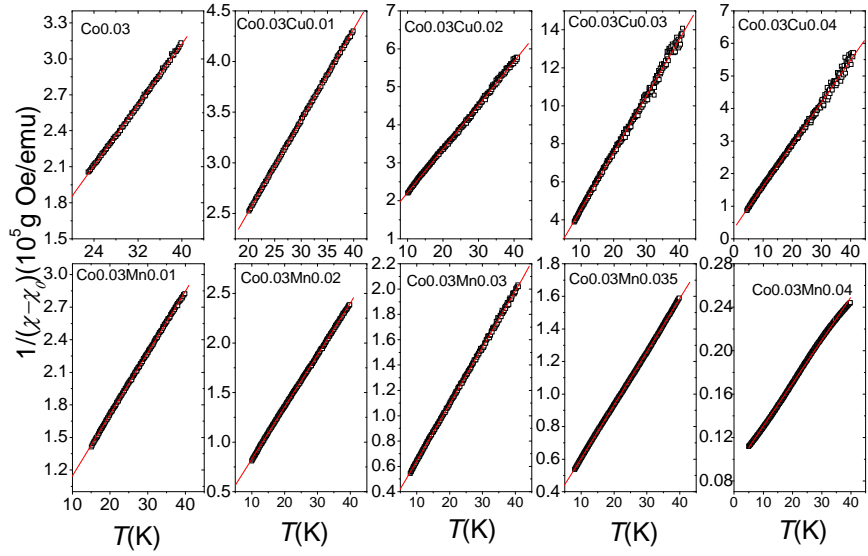

**Fig. S2.** Temperature dependence of DC magnetic susceptibility for Na(Fe$_{0.97}$Co$_{0.03}$)As and Na(Fe$_{0.96}$Co$_{0.03}$T$_{0.01}$)As (T = Cu and Mn) single crystals under 1 T. The $1/(\chi-\chi_0)$ is plotted as a linear function of $T$ in the low temperature limit. The red lines represent the linear fits of the data. The slope gives $1/C_0$, and the intercept provides the value of $T_\theta/C_0$.

### III. Illustration of the selective substitution of the Cu and Mn at Fe sites

In our previous paper, we demonstrated clearly that why each of the 3×2 block of the Na atoms on the surface corresponds to the doping site of Co. Here we use a cartoon picture, as shown in Fig. S3, to illustrate the selective substitution of Cu or Mn to the Fe sites, which leads to the two perpendicular orientations. These dumbbell shapes (in case of Cu dopants) align themselves along either [100] or [010] directions. This can be understood by considering the arrangement of Na atoms in one 3×2 block, we argue that each block is corresponding to one individual Cu impurity atom. Two strong reasons can be given to support this argument:

(1) When one Cu (the same for Co or Mn) atom is doped to the Fe site, it naturally neighbors with six top layer Na atoms, forming a 2-unit-cell rectangular (3×2) block (as illustrated by Fig. S3), the orientation of the 3×2 block is determined by the selective positioning of the Cu atom on the a-axis or b-axis of the Fe-As square lattice; (2) When counting the ratio between the number of the defects (brighter area) with this typical 3×2 block and the total Fe atoms, we find that it is very close to 1%, which is just the ratio of the chemical composition of Cu/(Fe+Co+Cu) in the sample. The same applies to the Mn dopants.

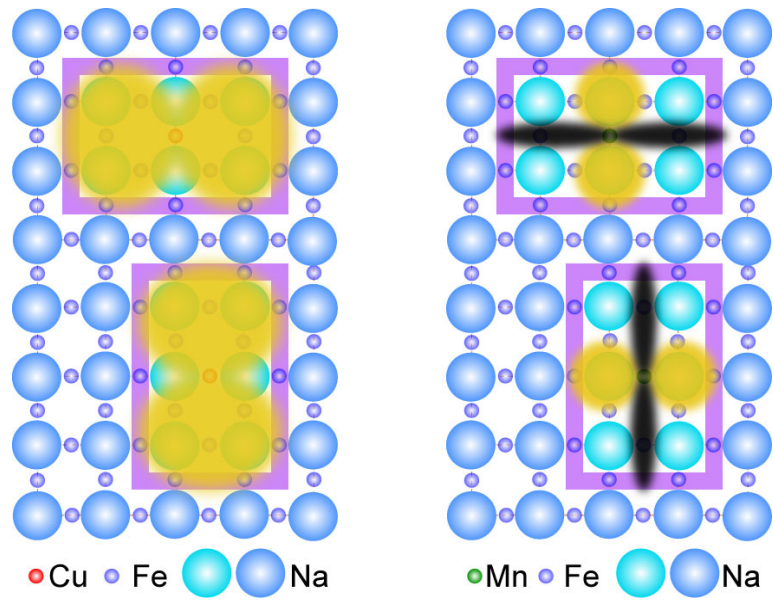

**Fig. S3.** Illustration of the selective substitution of Cu, Co and Mn to Fe sites. Left hand side shows the case for doping of Cu and Co. The right hand side shows the case for doping Mn. The yellow patches highlight the areas with enhanced brightness.

**IV. Scanning tunneling spectrum and the normalization**

The scanning tunneling spectra were measured with an ultrahigh vacuum, low temperature and high magnetic field scanning probe microscope USM-1300 (Unisoku Co., Ltd.). The samples were cleaved at room temperature in an ultrahigh vacuum with a base pressure better than $1.5\times10^{-10}$ torr. In all STM/STS measurements, Pt/Ir tips were used. To lower down the noise of the differential conductance spectra, a typical lock-in technique with an ac modulation at 987.5 Hz was used.

The STS measured at 1.7 K and 17 K at a Cu impurity and far away are shown in Fig. S4. The normalized curve presented in Fig. 5**A** is obtained by dividing the data measured at 1.7 K with that at 17 K, respectively.

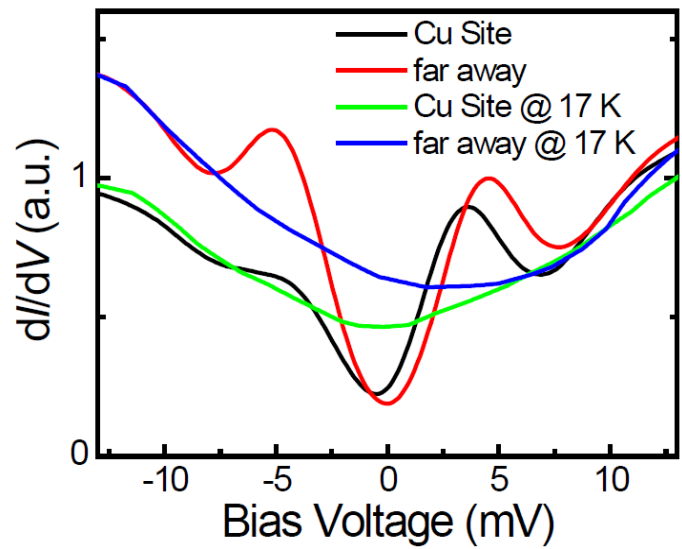

**Fig. S4.** Scanning tunneling spectra taken at 1.7 K and 17 K on the Na(Fe$_{0.96}$Co$_{0.03}$Cu$_{0.01}$)As single crystal on the Cu sites and far away. The spectra measured at 17 K were used as the background of the normal state.

**V. Theoretical calculations**

We used Quantum-Expresso (*1*) to get the band structure and subsequently the tight-binding model in terms of the maximally localized Wannier orbitals (*2*). The energy dispersion is shown in Fig.S5A along high symmetry cuts, and compares well to that in the literature (*3*). Fig. S5B shows the Fermi surface relevant to the material.

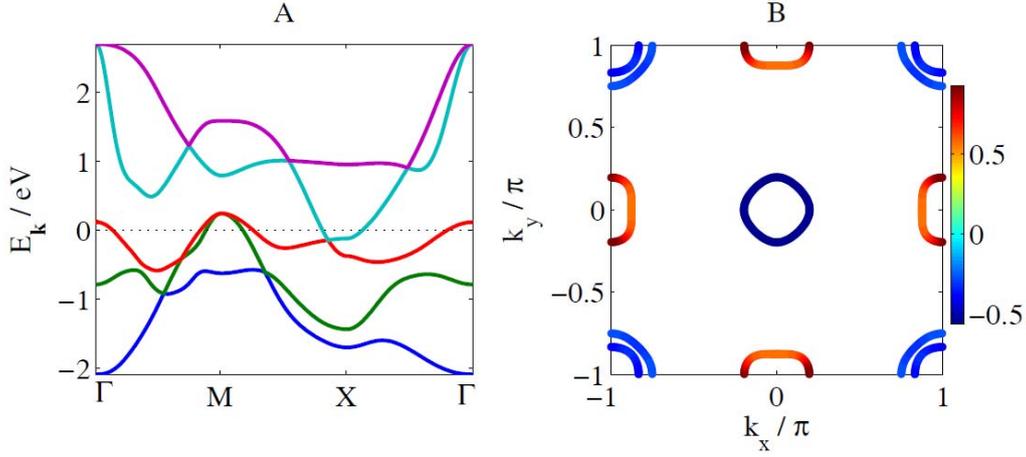

Fig. S5. (**A**) Band structure along high symmetry cuts. (**B**) Fermi surfaces. The color scale shows the pairing function projected on the Fermi surfaces.

For the *d*-orbital system under concern, it is sufficient to consider local interactions. These include intra-orbital repulsion *U*, inter-orbital repulsion *U'*, Hund's rule coupling *J* and pairing hopping *J'*. As usual we use the Murakami relation *U=U'+2J* and *J=J'*. For definiteness but without loss of generality we set *(U,J)=(1.00,0.30)*eV.

We use the singular-mode functional renormalized group (SMFRG) (*4-7*) to handle the effect of the interactions. It provides the coupled flow of the effective interaction vertex functions in the various density-wave and pairing channels versus a decreasing energy scale (say the temperature). The fastest growing eigenvalue of the vertex functions implies an emergent order associated with an ordering wave vector (which is zero in the Cooper pairing channel), and the corresponding eigenfunction dictates the internal structure of the order parameter. In our case, the pairing channel eventually dominates at low energy scales, but it is triggered by spin-density-wave fluctuations at moderate energy scales. The orbital-basis pairing function $\Delta_{ab}(k)$, where $a$ and $b$ label the five $d$-orbitals and $k$ is the momentum, is just the leading eigenfunction in the pairing channel. The pairing function projected on the Fermi surface is shown in Fig. S5B (color scale), revealing $S^{\pm}$ sign structure on the electron and hole pockets.

With the pairing instability predicted by the SMFRG, the ordered state can be reliably described by a corresponding effective BCS Hamiltonian, written in the Nambu space,

$$H_{BCS} = \sum_{a,b,k} \psi_{ak}^{+} (\varepsilon_k^{ab} \sigma_3 + \Delta_k^{ab} \sigma_1) \psi_{bk},$$

where $\varepsilon_k^{ab}$ is the tight-binding dispersion in orbital basis, $\sigma_{1,3}$ are Pauli matrices, and the global scale of the pairing function (on the Fermi surfaces) is tuned to the experimental value. In practice, the gap scale is tuned a few times larger in order to enhance the resolution in the numerical calculation of the density of states below, without changing the qualitative behavior.

We model the nonmagnetic impurity by an excess term in the Hamiltonian,

$$H_{imp} = \sum_a \psi_a^{+} V_{imp} \sigma_3 \psi_a,$$

at a local position in real space. The effect of such an impurity is conveniently treated by the T-matrix formalism. Let $G_0(i,j,z)$ denotes the unperturbed Green's function (a matrix in Nambu and orbital bases), where $i$ and $j$ are spatial positions and $z$ is a complex frequency. The impurity perturbed Green's function is given by

$$G(i,j,z) = G_0(i,j,z) + G_0(i,0,z) T(z) G_0(0,j,z),$$

where the T-matrix is given by

$$T^{-1}(z) = V_{imp}^{-1}\sigma_3 - G_0(0,0,z).$$

The local density of states at position $i$ is given by

$$\rho(i,\omega) = -\frac{1}{\pi}\text{Im}\left[\sum_a G_{aa}^{11}(i,i,z \to \omega + i\eta)\right],$$

where the numerical superscripts refer to the Nambu components. Here a Dynes smearing factor $\eta$ is used to take into account of the finite quasiparticle lifetime in experiments. In the presence of the impurity, new states within the pairing gap may be induced, and is given by the pole(s) of $T(z)$. For a one-band $s$-wave superconductor, it can be shown that the pole(s) can only appear at the gap edge for a nonmagnetic impurity, consistent with the Anderson theorem. For the $S^{\pm}$ wave case under concern, however, the quasi-particles see different signs of the pairing gap as they are scattered by the impurity between the bands. Such processes are known to induce in-gap states, in a similar manner to the magnetic impurity in a one-band $s$-wave superconductor, except that the in-gap states must be particle-hole asymmetric for a nonmagnetic impurity. We stress however that a zero-energy in-gap state only occurs at a specific value of the impurity potential, while nonzero-energy in-gap states are more generic. The theoretical result is shown in Fig. 5**C** and **D** of the main text. The impurity potential we used is $V_{imp}$ = -1.5 eV. Interestingly the induced in-gap peak is relatively stronger on the positive energy side. This result is related to the fact that the size of such a potential is of the order of the bandwidth, leading to a large spectral weight transfer to the lower band edge (not shown). This result has two important implications. First, it implies that the impurity is attractive, i.e., the number of electrons on the impurity is larger than elsewhere. Second, the unusual position of the in-gap peak can only be captured by the microscopic model with the realistic band structure. To see this, we also performed analytical calculations assuming wide band limit. The positive-energy peak for an attractive impurity appears only if the inter-band scattering is stronger than the intra-band scattering.